\documentclass[prb,twocolumn,showpacs,preprintnumbers,amsmath,amssymb,superscriptaddress, letterpaper]{revtex4}
\usepackage{graphicx}%
\usepackage{dcolumn}%
\usepackage{bm}%
\newcommand{\lco}{La$_2$CuO$_4$}
\newcommand{\ccoc}{Ca$_2$CuO$_2$Cl$_2$}
\newcommand{\hg}{HgBa$_2$CuO$_{4+\delta}$}
\newcommand{\K}{$K$}

\begin{document}

\title{Incident-energy and polarization-dependent resonant inelastic x-ray scattering study of \lco}

\author{L. Lu}
\affiliation{Department of Applied Physics, Stanford University,
Stanford, California 94305}

\author{J. N. Hancock}
\affiliation{Stanford Synchrotron Radiation Laboratory, Stanford,
California 94309}

\author{G. Chabot-Couture}
\affiliation{Department of Applied Physics, Stanford University,
Stanford, California 94305}

\author{K. Ishii}
\affiliation{Synchrotron Radiation Research Unit, Japan Atomic
Energy Agency, Hyogo 679-5148, Japan}

\author{O. P. Vajk}
\affiliation{NIST Center for Neutron Research, National Institute of
Standards and Technology, Gaithersburg, Maryland 20899}

\author{G. Yu}
\affiliation{Department of Physics, Stanford University, Stanford,
California 94305}

\author{J. Mizuki}
\affiliation{Synchrotron Radiation Research Unit, Japan Atomic
Energy Agency, Hyogo 679-5148, Japan}

\author{D. Casa}
\affiliation{CMC-XOR, Advanced Photon Source, Argonne National
Laboratory, Argonne, Illinois 60439}

\author{T. Gog}
\affiliation{CMC-XOR, Advanced Photon Source, Argonne National
Laboratory, Argonne, Illinois 60439}

\author{M. Greven}
\affiliation{Department of Applied Physics, Stanford University,
Stanford, California 94305} \affiliation{Stanford Synchrotron
Radiation Laboratory, Stanford, California 94309}

\date{\today}

\begin{abstract}

We present a detailed Cu $K$-edge resonant inelastic X-ray
scattering (RIXS) study of the Mott insulator \lco~in the 1-7 eV energy loss range. As initially
found for the high-temperature superconductor
\hg, the spectra exhibit a multiplet of weakly-dispersive
electron-hole excitations, which are revealed by utilizing the subtle
dependence of the cross section on the incident photon energy. The
close similarity between the fine structures for in-plane and
out-of-plane polarizations is indicative of the central role
played by the $1s$ core hole in inducing charge excitations within
the CuO$_2$ planes.
On the other hand, we observe a polarization dependence of the
spectral weight, and careful analysis reveals two separate features
near 2 eV that may be related to different
processes. The polarization dependence
indicates that the 4$p$ electrons contribute significantly to the
RIXS cross section. Third-order perturbation arguments and a shake-up of
valence excitations are then applied to account for the final-energy
resonance in the spectra. As an alternative scenario,
we discuss fluorescence-like emission
processes  due to $1s \rightarrow 4p$ transitions into a narrow
continuum $4p$ band.

\end{abstract}

\maketitle

\section{Introduction}
With the advent of third generation synchrotron sources, inelastic
X-ray scattering has emerged as a powerful probe of momentum- and
energy-dependent charge and lattice dynamics. This development has
led to new insight into low-density metallic electrodynamics
\cite{burns99}, valence fluctuating compounds \cite{dallera02},
H$_2$O molecular correlations \cite{wernet04,sette95}, phonon
dynamics \cite{dastuto02}, and the Mott physics of correlated
electron systems such as the lamellar copper oxides
\cite{hill98,abba99,hama00,hasan00,yjk02,yjk04,lu05,ishii05a,ishii05b}
and the manganites \cite{ishii04,grenier05}. Resonant inelastic
X-ray scattering (RIXS) provides a considerable advantage over
ordinary inelastic scattering since, at resonance, the inelastic
signal is significantly enhanced \cite{Kao96}. In the lamellar
copper oxides, this resonance condition can be readily met by tuning
the incoming photon energy to the vicinity of the Cu $K$ edge. The
RIXS cross section sensitively depends on the incident photon energy
and on the nature of the intermediate states
\cite{lu05,krisch95,doring04,uwe05}.

If viewed as a two-stage process, the intermediate state in RIXS is
the same as the final state of x-ray absorption: for example, in Cu
K-edge RIXS, a localized core hole is created through $1s
\rightarrow 4p$ photoexcitation. The $1s$ core hole interacts
strongly with the valence electron system, generating a strong
response that corresponds to the many-electron bound states of the
local, nascent core-hole potential.\cite{tranq91} In RIXS, the
relaxation of these highly excited states leads to the emission of
photons and leaves the valence system in an excited state. One
usually identifies energy-loss features with the excitations of the
valence electrons. When viewed as a second-order optical process,
there exists a close connection between the initial absorption and
final emission stages in RIXS. Accordingly, the spectra
simultaneously depend on both the incident and final photon
energies.\cite{kotani01}

In the present work, we investigate these energy dependences as well
as the polarization dependence of the cross section in
La$_2$CuO$_4$, the best-characterized lamellar copper oxide. This
Mott insulator is the parent compound of the original
high-temperature superconductor (La,Ba)$_2$CuO$_4$, and it has been
the subject of a number of prior RIXS
studies.\cite{yjk02,lu05,collart06} Exploiting the incident-energy
sensitivity, we are able to identify additional charge excitation
features. We demonstrate that the fine structure is present for
photon polarization both parallel and perpendicular to the CuO$_2$
planes, and suggest that the subtle differences between the two
polarization conditions can be explained in terms of models in which
the $4p$ electrons play a significant role.

This paper is organized as follows. After the discussion of the
experimental details in the next section, we present our results for
out-of-plane and in-plane polarization in Secs. III and IV,
respectively. Section V contains a discussion of our data, and we
summarize our work in Sec. VI.

\begin{figure}{h}
\begin{center}
\includegraphics[width=3.4in]{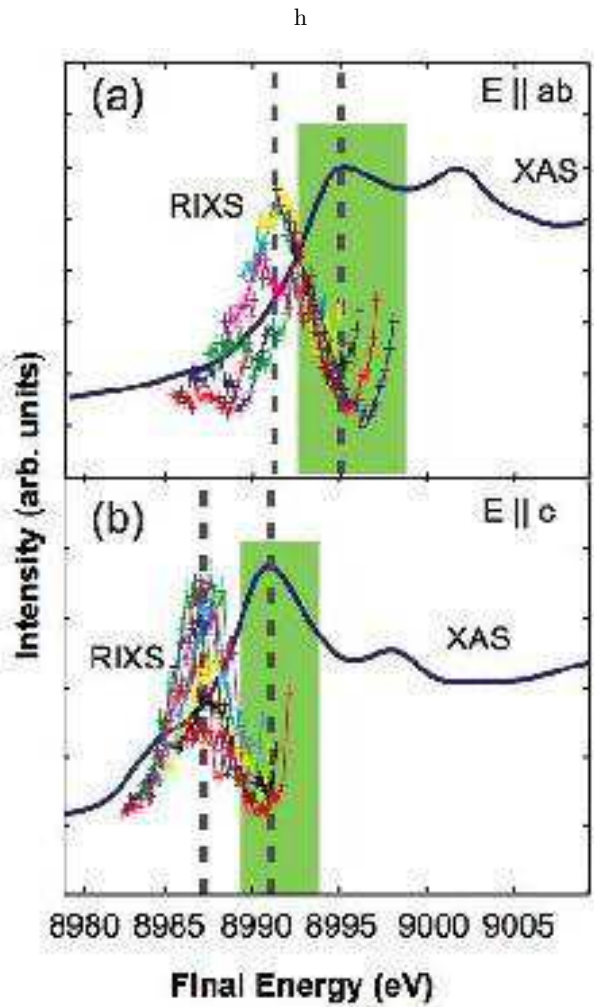}
\caption{(Color online) The two scattering geometries used in this
study and discussed in the text.} \label{fig:scatgeo}
\end{center}
\end{figure}
\section{Experimental Details}

The focus of this work is on the collective electronic excitations
of \lco\ in the 1–7 eV range using incident photon energies in the
vicinity of the Cu $K$-edge absorption threshold. We have measured
the RIXS response at several high-symmetry positions in the
Brillouin zone, covering a fine mesh of incident and scattered
photon energies around the Cu \K\ edge.

Two sets of measurements were taken, one collected at the Advanced
Photon Source with incident photon polarization vector perpendicular
to the CuO$_2$ planes (\textbf{E}$||c$), and the other at SPring-8
(Japan) with incident photon polarization parallel to the CuO$_2$
planes (\textbf{E}$\bot c$). The measurements with in-plane
polarization [Fig. \ref{fig:scatgeo} (a)] were taken in horizontal
scattering geometry at beamline BL11XU at SPring-8, with the
tetragonal reciprocal lattice point \textbf{G}=(3,0,0) as Brillouin
zone center, and a scattering angle of $\sim 65^{\circ}$. A Si(111)
main monochromator and a Si(400) secondary monochromator were used
to obtain an incident energy resolution of 220 meV. A bent Ge(733)
analyzer crystal situated at the end of a 2 m four-circle
diffractometer arm selected the energy of the photons scattered from
the sample, which were then collected by a solid state detector. In
this geometry, the polarization vector of the incident photon was
always parallel to the CuO$_2$ planes, with a typical angle of
$\sim32^{\circ}$ with respect to the tetragonal $a$ axis, i.e., the
planar Cu-O bond direction. The overall energy resolution was about
400 meV [full width at half maximum (FWHM)], as determined from the
energy width of the elastic line.

Out-of-plane polarization measurements [Fig. \ref{fig:scatgeo} (b)]
were performed at beamline 9-ID-B at the Advanced Photon Source in a
vertical scattering geometry. The reciprocal lattice vectors (3,0,0)
and (1,0,0) were chosen as reference zone centers to reduce the
contribution from the elastic tail, because Bragg scattering at
these reflections is forbidden. The setup employed a Si(111) primary
monochromator, a Si(333) secondary monochromator, and a spherical
diced Ge(733) analyzer crystal with a 1 m radius, and yielded an
overall energy resolution of about 300 meV (FWHM). For data
collected that were measured at the reduced wave vector ($\pi$,0),
or absolute momentum of  (1.5,0,0), we used a primary Si(111) and a
second Si(444) channel-cut monochromator in conjunction with a diced
analyzer on a 2 m diameter Rowland circle. This configuration can
provide at best an energy resolution of 110 meV, but we chose wide
aperture slits in front of the detector to obtain a significant
signal boost and a comparable resolution of $\sim$300 meV for better
comparison with measurements at other wave vectors.

Data were taken at ambient temperature on the same
single-crystalline sample in both polarization geometries.
\lco~undergoes a tetragonal-to-orthorhombic structural phase
transition at $ \sim 530$ K associated with the staggered tilting of
the CuO6 octahedra.\cite{BirgeneauPRB99} We note that our crystal is
twinned, and hence we do not distinguish between the two
inequivalent planar orthorhombic directions. The crystal was grown
in an image furnace at Stanford University. As-grown crystals are
known to contain excess oxygen, and hence hole carriers. In order to
assure that the sample was free of any carriers it was annealed for
24 h in Ar flow at 950$^{\circ}$C. This reduction treatment resulted
in a N\'eel temperature of $T_{\rm N}\sim320$ K, as determined from
a measurement of the magnetic susceptibility.\cite{vajk02}

\begin{figure}
\begin{center}
\includegraphics[width=2.8in]{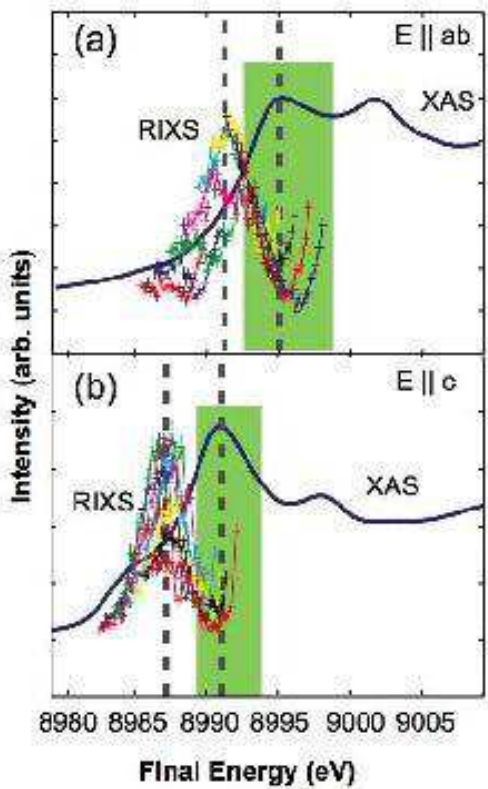}
\caption{(Color online) RIXS spectra plotted versus final photon energy along with
the absorption spectra monitored by total fluorescence yield for
photons polarized (a) parallel and (b) perpendicular to the CuO$_2$
planes. The shaded areas indicate the incident-energy ranges probed
in the present study. All spectra were taken at \textbf{Q} = (3,0,0) and
at different incident energies corresponding to the shaded ranges.} \label{fig:fluo}
\end{center}
\end{figure}

Figure \ref{fig:fluo} shows the x-ray absorption spectra (XAS) for
each polarization condition as measured by total fluorescence yield.
For each polarization, there are two peaks, at 8991 and 8998 eV for
\textbf{E}$||c$, and at 8995 and 9002 eV for \textbf{E}$\bot c$. The
lower of these resonances is usually identified with a transition
into a "well-screened state,"\cite{hill98,hama00,yjk02} a many-body
excitation which effectively screens the 1s core hole and has
significant $\underline{1s}4p3d^{10}\underline{L}$ character. The
higher resonance is identified with a transition into a "poorly
screened state,"\cite{hill98,hama00,yjk02} another bound state of
the many-electron system in the core hole potential which has
predominantly $\underline{1s}4p3d^{9}$ character. For each
polarization, the resonance peaks are separated by approximately 7
eV. The resonance energies differ by about 4 eV between the two
geometries, which may be primarily due to the larger Cu-O distance
for the negatively charged apical oxygens, rendering the 4$p_z$
electronic orbitals lower in energy than their 4$p_\sigma$
counterparts.\cite{heald88, lee06} The spectra presented in this
paper were taken in the vicinity of the well-screened condition for
both polarization geometries.

\section{Incident Energy Dependence, Out-of-plane polarization}

\begin{figure}
\begin{center}
\includegraphics[width=3.4in]{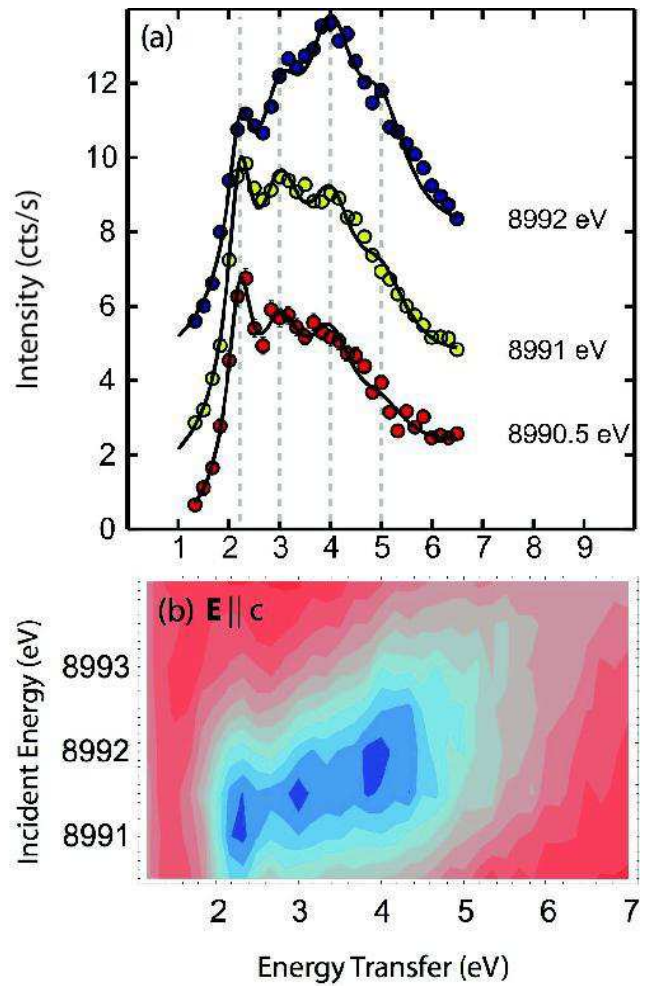}
\caption{(Color online) (a) RIXS signal versus energy transfer for out-of-plane
polarization at three representative incident energies at a momentum
transfer of (3,0,0), which corresponds to the
two-dimensional zone center (0,0). The lines are the result of a
fit, as discussed in the text (reproduced from \cite{lu05}).
(b) Contour plot of all zone-center scans, taken in 500 meV
increments of $E_{\rm i}$.} \label{fig:cpol}
\end{center}
\end{figure}

RIXS spectra obtained in early work in the soft x-ray regime
exhibited a clear incident and final photon energy
dependence.\cite{kotani01} The incident-photon-energy dependence was
recently employed in the hard x-ray regime (at the Cu $K$ edge) in a
study of both \lco~and the single-layer hightemperature
superconductor \hg~(Hg1201).\cite{lu05} This study revealed
additional features in the 2-5 eV range, an observation that
necessitates a new interpretation of the charge dynamics in these
materials. For example, a $\sim 2$ eV feature was identified in
Hg1201. It was argued in Ref. \onlinecite{lu05} that this feature is
not likely a $d \rightarrow d$ excitation, but rather indicates the
presence of a remnant charge-transfer gap even at optimal doping in
this model superconductor. The presence of an additional feature at
$\sim$3 eV, which was only identified through inspection of multiple
spectra obtained with different photon energies, constrains the
dispersion of the 2 eV feature to be less than 500 meV. The same
approach was applied to \lco, and preliminary data revealed
charge-transfer features that are remarkably similar to those in
Hg1201,\cite{lu05} a result that is qualitatively different from
prior work on the Mott insulators \lco~\cite{yjk02} and
\ccoc~\cite{hasan00} The small dispersion of the 2 eV feature was
further confirmed by subsequent measurements.\cite{collart06, kim06}
Our primary focus here is to investigate in greater detail the
incident-energy and polarization dependence of the inelastic cross
section near the absorption threshold in \lco.

The molecular orbital excitation at $\sim 7$ eV was studied in
detail in Ref. \onlinecite{yjk04} and is most prominently observed
near $E_i$ = 8998 eV, an incident photon energy for which the
lower-lying charge-transfer excitations in the 2-6 eV range do not
resonate. We therefore limit our attention to the incident energy
range $E_i$ = 8989-8994 eV [shaded area in Fig. \ref{fig:fluo} (a)]
and to energy transfers below 7 eV for out-of-plane polarization. A
fine step size of $\Delta E_i = 500$ meV was chosen, allowing us to
demonstrate the high sensitivity of the RIXS cross section to the
incident photon energy.

Figure \ref{fig:cpol} (a) shows representative line scans at the
zone center, taken at three different incident energies with
out-of-plane polarization. These data resemble previous work
\cite{yjk02}, yet closer inspection reveals additional features. At
low incident energy, the most distinct feature is that at 2.25 eV.
As the incident energy is increased, this sharp feature gradually
weakens relative to those at higher energy.  Another new feature at
3 eV, which was not observed in prior work \cite{yjk02}, can be
discerned at all three incident photon energies. The feature at
$\sim 4$ eV, also seen in previous zone-center data \cite{yjk02},
becomes dominant at $E_i = 8992$ eV. Finally, a comprehensive
analysis of all data \cite{lu05} reveals a second new feature at
$\sim 5$ eV. We discuss below how the systematic center-of-mass
shift with incident energy suggests a modulation of the inelastic
cross section through final photon-energy-dependent denominators.

Figure \ref{fig:cpol} (b) shows a contour plot constructed from all
line scans at the zone center. This mode of representation is
similar to the ``RIXS plane" of incident photon energy versus energy
transfer in Ref. \onlinecite{glatzel05}. By extending the
energy-transfer spectra into the incident-energy dimension, features
at $\sim 2.25$, 3 and 4 eV are readily apparent.

\begin{figure}
\includegraphics[width=3.4in]{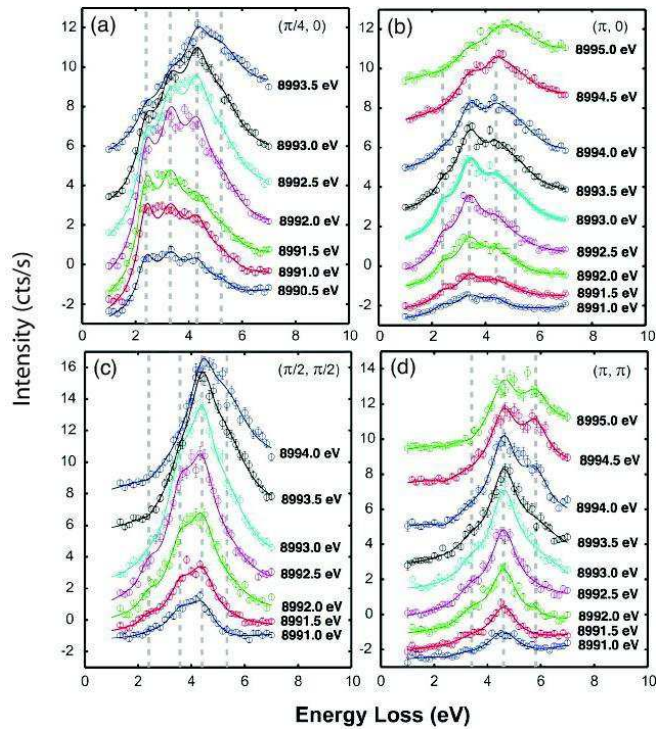}
\begin{center}
\caption{(Color online) Incident-energy dependence of the \textbf{E}$||c$ RIXS
spectra for (a) ($\pi$/4,0), (b) ($\pi$,0), (c) ($\pi/2$,$\pi/2$),
(d) ($\pi$,$\pi$). The lines are the result of fits, as discussed in
the text.} \label{fig:cdisp}
\end{center}
\end{figure}

\begin{figure}
\begin{center}
\includegraphics[width=2.6in]{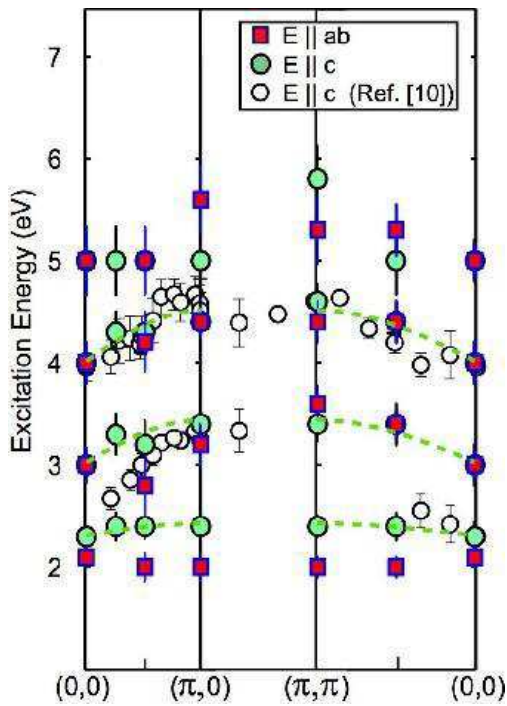}
\caption{(Color online) Dispersion of the charge excitations for both in-plane and
out-of-plane polarization along the two high-symmetry directions studied
in the present work. The dashed lines indicate sinusoidal fits to the
dispersion of the three lowest-energy features discerned with out-of-plane
polarization. Previous results \cite{yjk02}, obtained with out-of-plane polarization,
are shown for comparison.
} \label{fig:dispersion}
\end{center}
\end{figure}

Figure \ref{fig:cdisp} shows additional results at high-symmetry
points of the two-dimensional Brillouin zone. Using the fit
procedure defined in Ref. \onlinecite{lu05}, the spectra at each
momentum transfer are fitted simultaneously to obtain the peak
positions that determine the energy transfer of the corresponding
charge excitations. Specifically, we assume that the peak positions
do not vary with incident energy and that each energy transfer
feature is represented by a Lorentzian line shape. The number,
energy-transfer positions, and energy widths of the features are
considered to be shared parameters for all spectra at the same
momentum transfer. At different incident energies, on the other
hand, the spectral weight of each component is allowed to vary. We
also use spectra on the energy gain side (not shown in Fig.
\ref{fig:cdisp}) for background subtraction, and allow for linear
slopes to approximate the continuum due to transitions to continuous
unoccupied states. A simultaneous least-squares fit of all spectra
at each momentum transfer results in the lines in Figs.
\ref{fig:cpol} (a) and \ref{fig:cdisp} and allowed us to extract the
peak positions plotted in Fig. \ref{fig:dispersion}.

Although the different components are not as easily distinguishable
as at the zone center, the relative strengths of the four features
identified below 6 eV exhibit distinct dependences on momentum
transfer and incident energy, which may be indicative of different
excitation symmetry. The 5 eV component is most pronounced at $(\pi,
\pi)$ and at relatively high incident energies. Away from the zone
center, the relative weight of the 2 eV feature quickly decreases.
While the 2 and 3 eV features are still separable and comparable in
strength at $(\pi/4, 0)$, the 2 eV component is only barely visible
(at low incident energies) at $(\pi,0)$, the 3 and 4 eV features
maintain comparable weight along $[\pi,0]$. In contrast, along
$[\pi,\pi]$, the overall response away from the zone center appears
to be dominated by the (approximately Lorentzian-shaped) response
just above 4 eV. As for the fine structure, the momentum dependence
away from the zone center is again consistent with what is observed
in Hg1201, although that study was carried out with in-plane
polarization.\cite{lu05}

\section{Incident Energy Dependence, in-plane polarization}

\begin{figure}
\begin{center}
\includegraphics[width=3.4in]{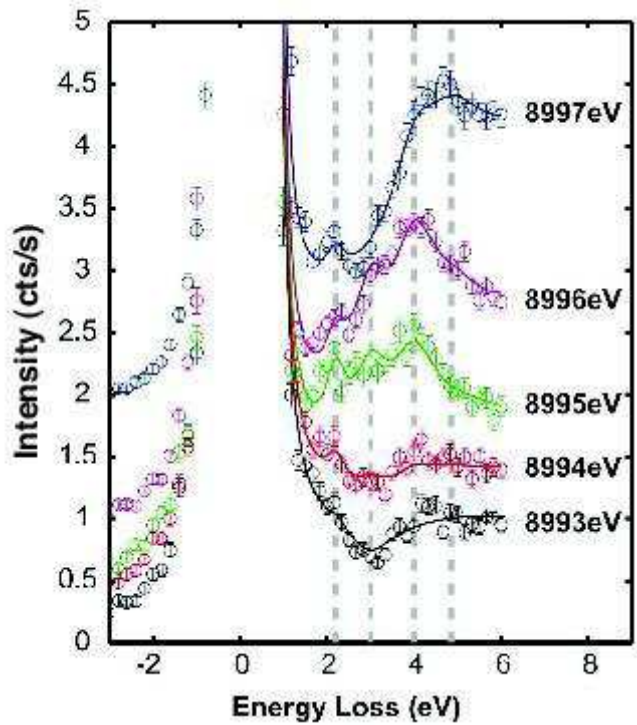}
\caption{(Color online) Five line scans at the zone center obtained with in-plane
polarization.}
\label{fig:abpol}
\end{center}
\end{figure}

Figure \ref{fig:abpol} shows line scans at the zone center for the
complementary in-plane-polarized (SPring-8) experiment. As indicated
in Fig. \ref{fig:fluo} (b), the incident photon energy was chosen to
lie in the range 8993 – 8999 eV. The fine structure revealed through
the incident energy dependence of the spectra is very similar to
that found for out-of-plane polarization. Differences in the
configuration between the two experiments make a direct comparison
of signal levels difficult. Overall, there are still four major
features present at $\sim 2$, 3, 4 and 5 eV. The close similarity of
the spectra supports the intuitive notion that the
spherically-symmetric $1s$ core hole potential dominates the
generation of the valence excitations in both
cases.\cite{mahan00,degroot01,nomura05,hasan00,tsu03}

\begin{figure}
\begin{center}
\includegraphics[width=3.4 in]{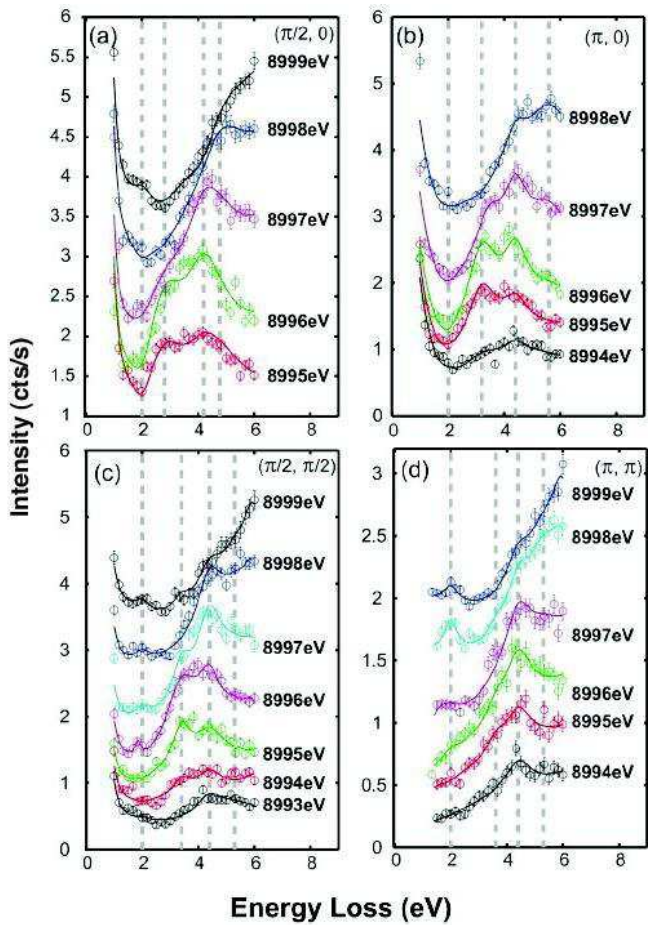}
\caption{(Color online) RIXS spectra with \textbf{E}$||ab$ at (a) ($\pi$/2,0)
(b) ($\pi$,0) (c) ($\pi/2$,$\pi/2$), and (d) ($\pi$ $\pi$).
}
\label{fig:abdisp}
\end{center}
\end{figure}

There also exist similarities between the two polarization
geometries in the momentum dependence of the multiplet structure.
Figure \ref{fig:abdisp} shows line scans with fits at four reduced
momentum transfer values away from the zone center. As in the
out-of-plane polarized experiment, the strength of the 2 eV feature
decreases toward the zone boundary. In the present case, it is no
longer visually observable at $(\pi, 0)$. Also, the 3 eV and 4 eV features
have comparable spectral weight 
along $[\pi,0]$, and remain comparable up to $(\pi/2, \pi/2)$ along
$[\pi, \pi]$. Between ($\pi$/2,$\pi/2$) and ($\pi$,$\pi$), however,
the 3 eV feature is quickly suppressed between and, as in the case
for out-of-plane polarization, the 4 eV spectral weight becomes
dominant.

Differences between the spectra obtained in the two geometries are
also noticeable. First, with in-plane polarization, the intensity of
the 2 eV feature relative to that of spectral features with higher
energy transfer is much smaller than for the out-of-plane polarized
experiment. However, the 2 eV feature is still observable even up to
$(\pi,\pi)$, reaching a maximum at an incident energy of 8997 eV,
slightly above the absorption threshold (8995 eV) for in-plane
polarization. Second, for in-plane polarization, as we increase in
incident energy, the center of mass of the spectra continuously
shifts to higher energy transfers. For out-of-plane polarization, on
the other hand, it peaks between 4 and 5 eV.

As for the {\bf E}$||c$ data, all spectra acquired at the same
momentum transfer were simultaneously fit assuming the same set of
peak positions. The results of the fits are shown by the lines in
Figs. \ref{fig:abpol} and \ref{fig:abdisp} and the peak positions
are compared to those for out-of-plane polarization in Fig.
\ref{fig:dispersion}. While there is an overall good agreement, the
2 eV features have significantly different excitation energies for
the two polarization conditions. We note that the dispersions of all
four features summarized in Fig. \ref{fig:dispersion} are weak.
These observations will be discussed in more detail in the next
section.

\section{Discussion}

The presence of a $\sim$ 2 eV resonance feature has been discussed
in connection with excitations across the charge transfer
gap.\cite{kotani01,hasan00,abba99, yjk02,lu05} In \lco, this
excitation is observed only for transitions into well-screened
states, in which the $1s$ core hole is screened by a valence
electron from a neighboring CuO$_4$ plaquette, leaving a doubly
occupied Cu$^+$ ion ($3d^{10}$) and a hole on the neighboring
plaquette. It has been suggested that the nonlocal hole can form a
Zhang-Rice singlet,\cite{zhang88} which can propagate efficiently
through the antiferromagnetic background, and that this singlet
could form a strong bond with the Cu$^+$ quasiparticle and become
even more dispersive as a bound exciton.\cite{zhang98} However,
high-resolution electron-energy-loss spectroscopy (EELS) on the
related Mott insulator Sr$_2$CuO$_2$Cl$_2$ suggests the existence of
another charge-transfer excitation, with slightly lower energy, that
involves only the local CuO$_4$ plaquette. These latter findings
were argued to be consistent with embedded molecular cluster
calculations.\cite{moskvin02} Indeed, our data reveal that the
actual excitation energies of the 2 meV feature differ by as much as
300 meV for in-plane and out-of-plane polarization conditions. While
the former excitation has no discernible dispersion, the latter
appears to disperse by 100 – 150 meV toward the zone boundary. In
addition to these differences, we also find that the spectral
weights of these two low-energy features exhibit rather different
momentum dependences, especially along $[\pi,\pi]$. We note that it
is not likely that the 2 eV features are $d\rightarrow d$
excitations, since the latter lie below 2 eV and are expected to be
much weaker at the $K$ edge than at the $L$ and $M$
edges.\cite{kotani01}

\begin{figure}
\begin{center}
\includegraphics[width=3.4in]{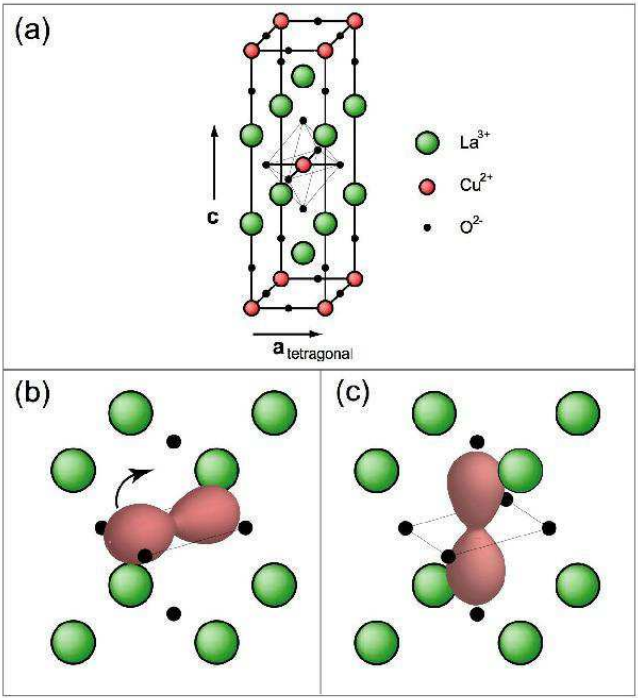}
\caption{(Color online) (a) \lco~unit cell. In a Cu $K$-edge absorption process,
several 4$p$ states can be reached, depending on the polarization of
the incident photon relative to the crystal axes. (b) 4$p_\sigma$
configuration created by absorption of a $E_{\rm i} \sim8996$ eV
photon with $E||ab$. (c) 4$p_z$ configuration created by
absorption of a $E_{\rm i}\sim8991$ eV photon with
$E||c$. The emissive process discussed in the text
corresponds to relaxation from state (b) to state (c).}
\label{fig:atoms}
\end{center}
\end{figure}

In an attempt to understand the differences between the two 2 eV
features, we consider a possible photon polarization effect. When
the polarization vector of the incident photon lies within the
CuO$_2$ plane, the $4p$ electron in the intermediate state is in the
$4p_\sigma$ orbital [see Fig. \ref{fig:atoms} (b)] and overlaps with
the $2p_{x,y}$ electrons. The repulsive Coulomb interaction between
O $2p$ and Cu $4p$ therefore tends to suppress the O$2p \Rightarrow
$Cu$3d$ charge transfer. On the other hand, for out-ofplane
polarization, the $4p_z$ orbital is oriented orthogonally to the
$2p_{x,y}$ electrons [see Fig. \ref{fig:atoms} (c)] and the Coulomb
interaction has a limited effect on the charge-transfer process.
That difference in intermediate states may explain why we observe a
second (local) ``2eV" component for in-plane polarization, since the
component that involves a nonlocal charge-transfer process from a
neighboring CuO$_4$ plaquette to the central core hole site could be
spectroscopically suppressed for in-plane polarization. The lower of
the two 2 eV features may be intrinsically weaker, which could
explain why we are unable to discern it for out-of-plane
polarization. As an alternative scenario, consistent with the
identification of two distinct features, a detailed analysis of the
scattering configurations reveals that the optical-limit Raman
efficiencies set by geometry are quite different in each case. It is
possible that the suppression of the higher energy feature for the
plane-polarized experiment reflects details of the symmetry of this
excitation. Further experimentation is required to resolve this
possibility. The above is consistent with earlier suggestions based
on Cu K-edge XAS.\cite{tolentino92} Especially for in-plane
polarization, the $\sim$2 eV excitation still maintains its strength
along [$\pi,\pi$], whereas it is weakened at the same location for
out-of-plane polarization. Therefore, it indeed appears that these
two low-energy excitations have a distinct physical origin,
consistent with the high-resolution EELS work.\cite{moskvin02}

The magnitude of the dispersion will be an important factor in the
eventual determination of the origin of the charge excitations. Our
\lco~data reveal charge-transfer features that are remarkably
similar to those for Hg1201,\cite{lu05} a result that is
qualitatively different from prior work on \lco .\cite{yjk02} As
summarized in Fig. \ref{fig:dispersion}, below 4 eV we identify one
additional branch at $\sim$3 eV for both polarization conditions.
For example, for out-of-plane polarization, simple fits to a
sinusoidal form (shown in Fig. 5) yield dispersions of 120(30),
410(110), and 490(70) meV for the 2, 3, and 4 eV features along both
high-symmetry directions. The 5 eV feature is dispersion-less within
the experimental uncertainty, and we note a possible anomaly at
($\pi$, $\pi$). Below, we discuss the possibility that the 5 eV
feature may actually be the result of a "shake-up" excitation at 7.2
eV. The $\sim$ 100 meV dispersion of the nonlocal 2 eV excitation is
less than the Zhang-Rice singlet bandwidth of 250 meV identified by
angle–resolved photoemission spectroscopy.\cite{wells95, durr00,
ronning03} This observation challenges the notion of an excitonic
picture to explain the dispersion of the charge-transfer gap
excitation. In principle, if the electronic states of the electron
and hole are asymmetric,\cite{veenendaal04} the observed dispersion
may either represent the bandwidth of the upper Hubbard band, if the
electron is more mobile, or of the Zhang-Rice singlet band, if the
hole is freer to move. However, it is difficult to reconcile the
small electron-hole-pair dispersion of $\sim$ 100 meV with the
relatively large Zhang-Rice singlet bandwidth, unless the observed
behavior represents a significant core hole effect.

In order to understand the appearance of multiple charge excitations
in the higher-energy (transfer) region, a more complex approach
appears to be necessary. An initial suggestion concerning the 4 eV
excitation invoked an excitonic state of unspecified
origin,\cite{yjk02} yet more recent work\cite{nomura05,mark05,lu05}
suggests that the higher-energy spectral features ought to be
described in terms of a multiband picture. Considering the
involvement of bonding and nonbonding oxygen $2p_\sigma$ and $2p_z$
orbitals, as well as charge-transfer processes through local and
nonlocal screening channels, there exist many candidate modes. Eskes
and Sawatzky\cite{eskes91} also find that triple-band physics,
including the Zhang-Rice triplet states, as well as
$d_{3z^2-r^2}$-orbitals are relevant up to about 7 eV in binding
energy. Further experimentation, including symmetry analysis, is
required to resolve the physical origin of the high-energy spectral
features.

\begin{figure}
\includegraphics[width=3.4in]{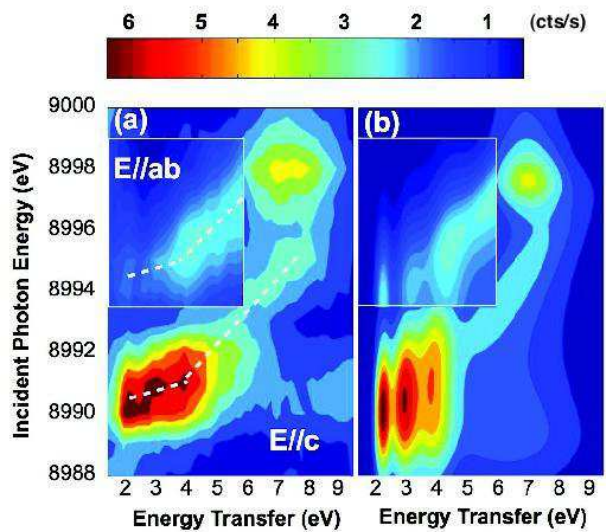}
\caption{(Color online) (a) Contour plot of \lco~RIXS intensity: incident photon
energy vs. energy transfer for E $||$ c (from Ref. \cite{yjk02}) and
E $||$ ab (inset; present work). The white dashed lines indicate the
two trends with different slope, as discussed in the text. (b)
Calculations following third-order perturbation theory, as discussed
in the text.
}
\label{fig:contour2}
\end{figure}

We will now discuss the photon energy and polarization dependence of
the observed higher-energy features. Figure \ref{fig:contour2} (a)
shows zone-center contour plots of incident energy versus energy
transfer. The out-of-plane data are from Ref. \onlinecite{yjk02} and
were taken with coarser incident energy step size (1 eV vs. 0.5 eV),
but span a wider incident energy range than our data in Fig.
\ref{fig:cpol} (b). The in-plane data (inset) are from our present
study and were taken with incident-energy increments of 1 eV. As
mentioned above, the center of mass of the RIXS spectra shifts to
higher energy transfer as the incident photon energy increases. This
variation is identified in Fig. \ref{fig:contour2} (a) as a
``streak" of intensity from 2 to $\sim$7 eV which, instead of
extending horizontally as one would expect for resonances associated
with fixed incident photon energy, tilts toward the upper right
corner. We will first discuss that this streak of intensity can be
interpreted in a shakeup picture, which uses third-order
perturbation theory. As an alternative scenario for the slope-1
component of this streak of intensity between $\sim$5 and 7 eV, we
will then discuss fluorescence-like emission processes due to $1s
\rightarrow 4p$ transitions into a narrow continuum $4p$ band.

Pioneering work on \lco~ (Ref. \onlinecite{abba99}) utilized
in-plane polarization and revealed one single excitation between
$\sim$3 and 6 eV, and it was found that the peak position varied
nonlinearly with incident energy. These results were interpreted in
terms of a shake-up of the $3d$ electron system, and explained
within third-order perturbation theory. Following this treatment,
which was formulated in detail in Ref. \onlinecite{abba99,
doring04,platzman98}, the scattering amplitude in third-order
perturbation theory is given by

\begin{eqnarray}
A &\propto& \sum_{m,n}\frac{\langle f|b_2| m\rangle \langle m | H_c
| n \rangle \langle n |
b_1|i\rangle}{(E_m-E_{f,el}-E_f-i\Gamma_m)(E_n-E_{i,el}-E_i-i\Gamma_n)}
\nonumber\\
&\approx& \frac{\langle f|b_2 H_c b_1|i\rangle}{(E_{ex,2}-E_f-i\Gamma_2)(E_{ex,1}-E_i-i\Gamma_1)}
\end{eqnarray}
where $|m\rangle$ and $|n\rangle$ denote different intermediate
states containing a virtual $\underline{1s}4p$ exciton, $E_i$ and $E_f$ are the incident and final
photon energies, $E_{i,el}$ and $E_{f,el}$
the initial energy and final energies of the electron system, and $b_1$ and $b_2$ are
absorption and emission operators \cite{doring04}. A simplification
is made by defining the constants $E_{ex,1} \equiv E_n - E_{i,el}$, and $E_{ex,2} \equiv E_m - E_{f,el}$, which can be considered incident and final resonant energies, respectively, and
the respective inverse lifetimes $\Gamma_1$ and $\Gamma_2$.
\cite{abba99,doring04}.
With this simplification, we ignore the details of the intermediate states.
We note that this formula contains separate
denominators involving incident and final photon energies.
The scattering intensity is given by
\begin{eqnarray}
I \sim &&L(E_{ex,1}-E_i, \Gamma_1)L(E_{ex,2} -E_f, \Gamma_2)
\nonumber\\
     &&\times  |\langle f|b_2H_cb_1|i\rangle|^2 \delta(E_{f,el} - E_{i,el} -  \omega),
\end{eqnarray}
where $L(E,\Gamma_i)$ represents a Lorentzian function with half width
$\Gamma_i$.

To model the shake-up process, we replace  $|\langle
f|b_2H_cb_1|i\rangle|^2$ with a sum of Gaussian
functions, each representing a distinct symmetry-allowed shake-up
excitation with energy $\Delta$ and heuristic inverse life-time
$\Gamma_s$:

\begin{equation}
G(\omega) = exp\left(-\frac{(\omega-\Delta)^2}{2\Gamma^2_s}\right).
\end{equation}

The single excitation observed in prior work on \lco~(Ref.
\onlinecite{abba99}) was described with $\Gamma_1 = \Gamma_2 = 2.38$
eV, $E_{ex,1} = E_{ex,2} = 8995$ eV, $ \Gamma_s  = 3.9$ eV and
$\Delta = 6.1$ eV. Since we have been able to resolve a multiplet of
excitations rather than a single excitation, we apply third-order
perturbation theory to the entire multiplet. In our calculation, we
find that four valence excitations with energies $\Delta = 2.3$, 3,
4  and 7.2 eV are adequate to represent the spectra for both
polarization conditions. The result of this calculation is shown in
Fig. \ref{fig:contour2} (b). For this particular calculation,
$\Gamma_s$ to be 0.4, 1.0, and 1.1 eV for the lower three
excitations, respectively. We find that our data can be adequately
represented if the 7.2 eV molecular orbital excitation resonates for
transitions into both well- and poorly screened states, with
variable characteristics. This excitation is represented by two
Gaussians (with $\Gamma_s$ set to 3.9 eV (well-screened) and 1.9 eV
(poorly screened)), and we therefore consider a total of five
Gaussians for four excitations.

We note that the 5 eV feature discussed in Secs. III and IV was not
considered in the above calculations since it is most prominent at
high incident energies and shifts with $E_i$. It therefore seems
likely that this feature may actually be associated with a resonance
of the 7.2 eV molecular orbital excitation at the well-screened
state. Due to the doubleresonance denominator, each component may,
in principle, show two separate resonances for $E_i = E_{ex,1}$ and
$E_f = E_{ex,2}$. For the final-energy resonance, as the incident
energy changes, the peak position of the excitation shifts to
$\omega_{peak} = E_i - E_{ex,2}$ so as to maintain the same final
energy. In the twodimensional contour plot of Fig.
\ref{fig:contour2}, this manifests itself as a slope-1 streak of
intensity.

The inverse lifetimes determine the shapes of the resonant spectral
features. When $\Gamma_1$, $\Gamma_2$, and $\Gamma_s$ approach the
value of excitation energy $\Delta$, the two resonances merge and
become indistinguishable. If $\Gamma_1$ or $\Gamma_2$ is small
compared to $\Delta$ and $\Gamma_s$, one of the two resonances
nearly disappears, and the response either is either a circular
region or a slope-one streak. By carefully choosing $\Gamma_i$ and
$E_{ex,i}$ ($i=1,2$) for each component, we we were able to emulate
the main characteristics of the experimental data. As seen from Fig.
\ref{fig:contour2} (a), the lowest three excitations are only
strongly resonant at the well-screened state, while the 7.2 eV
excitation is associated with the poorly-screened state.
Consequently, $E_{ex}$ should be different for the latter. For the
slope-1 streak component, the incident resonance energy is the same
as for the lower three excitations, but we find it necessary to
choose a slightly smaller $E_{ex,2}$.

For out-of-plane polarization, we set $\Gamma_1$ = $\Gamma_2$ = 2 eV
and $E_{ex,1}$=$E_{ex,2}$ = 8990 eV for the lower three excitations.
For the 7.2 eV component that results in the diagonal streak, we set
$\Gamma_1$ = 1.7 eV, $\Gamma_2$ = 1.5 eV,  $E_{ex,1}$= 8990 eV, and
$E_{ex,2}$ = 8989 eV, and for the other 7.2 eV component we chose
$\Gamma_1$ = 1.0 eV, $\Gamma_2$ = 1.5 eV and $E_{ex,1}$= $E_{ex,2}$
= 8998.5 eV. For in-plane polarization, we simply shifted all
$E_{ex}$ values by 4 eV,  and adjusted the relative intensity
between the three low-lying excitations and the two 7.2 eV
components.

The above analysis has two important implications. One is that the
``5 eV" feature is to be viewed as a shake-up excitation at 7.2 eV.
The second important implication is that this local
molecular-orbital excitation not only resonates at the poorly
screened state at which copper has an open-shell configuration
($\underline{1s}3d^94p$), but also at the well-screened state
($\underline{1s}3d^{10}\underline{L}4p$). This is different from
findings for CuO and from the argument that the observed shake-up
excitation requires the $3d^9$ open-shell configuration and should
be absent at the well-screened state \cite{doring04}. However, we
note that the molecular-orbital excitation was observed to be
resonant at both well-screened and poorly-screened states in
superconducting \hg~\cite{lu05}.


We will now discuss a different interpretation for the slope-1
component of the streak of intensity that relates it to a resonant
excitation into the continuum of unoccupied states.\cite{gel98,
gel99} Figure \ref{fig:fluo} shows the zone-center RIXS spectra
together with the respective absorption spectra for both
polarization condtions. The RIXS intensity is plotted versus final
photon energy ($E_f$) instead of energy transfer ($\omega$). We find
that the spectra collapse to a peak, with an envelope of
approximately Lorentzian shape, centered at about $E_f=8991$ eV for
in-plane polarization and $E_f=8987$ eV for out-of-plane
polarization. The peak position lies $\sim$ 4 eV below the
photoabsorption threshold for both polarization conditions. This
observation is consistent with the presence of slope-1 streaks in
the contour plots for both polarization conditions: for in-plane
polarization this streak starts at $E_i$ $\sim$ 8995 eV and for
out-of-plane polarization it starts at $E_i$ $\sim$ 8991 eV.

The photon-absorption process near the Cu $K$ edge is comprised of
transitions to either narrow molecular orbitals or continuum
unoccupied states. For discrete levels, such as the well-screened
and poorly screened states, induced valence excitations resonate and
follow the Raman-Stokes law as $E_i$ crosses the discrete levels,
i.e., the excitation energies $\Delta$ do not vary with incident
photon energy. On the other hand, a resonance due to transitions to
continuum states behaves differently. Here, the resonance condition
is fulfilled for every incident energy tuned to the continuum, and
the subsequent emission is independent of $E_i$ once the incident
energy increases above the lower edge of the continuum. When $E_i$
is below the edge, the resonant emission should also follow the
Raman-Stokes behavior, but the spectral weight may be suppressed due
to the small density of states below the edge. Our observation is
consistent with the existence of a continuous unoccupied band of
$4p$ symmetry, with an edge at $\sim$ 8991 eV and $\sim$ 8995 eV for
the respective polarizations, and a width of about 3–4 eV. This
would support the view that a combination of an extended picture of
itinerant $4p$ electrons and of localized molecular orbitals is
necessary to interpret the $K$-edge absorption spectra\cite{tranq91}
and the nature of the intermediate states in RIXS.

We note that the envelope of the in-plane data collapse shown in
Fig. \ref{fig:fluo}(a) is very similar in shape and position to the
XAS corresponding to the well-screened state for out-of-plane
polarization [Fig. \ref{fig:fluo}(b)]. This leads us to a third
interpretation of the slope-one contribution to the RIXS cross
section shown in Fig. \ref{fig:contour2}(a). Specifically, it
suggests that part of the in-plane-polarized RIXS signal can be
interpreted through the following complex dynamical process
illustrated in Fig. \ref{fig:atoms}: \lco absorbs a photon of energy
$E_i\sim$ 8995 eV with polarization \textbf{E}$\bot c$, creating on
a Cu site a well-screened $1s$ core hole and an electron in a
$4p_\sigma$ orbital [Fig. \ref{fig:atoms} (b)]. The $4p_\sigma$
electron then evolves into a $4p_z$ state [Fig. \ref{fig:atoms} (c)]
via a subsequent relaxation process. Finally, the 4pz electron
recombines with the $1s$ core hole, emitting a photon with energy
$\sim$ 8991 eV. However, for out-of-plane polarization, electrons
are already excited into a $4p_z$ state, and this relaxation will
not occur. Nevertheless, the envelope for out-of-plane polarization
lies also 4 eV below that of the main edge. This interpretation
therefore requires the existence of a discrete lower-energy state
with energy 8987 eV for 4pz electron to relax into. Interestingly,
in Ref. \onlinecite{abba99}, a resonance of approximately this
energy was observed with in-plane polarization. Further
experimentation combined with theoretical modeling can be expected
to resolve the origin of the slope-1 streak of intensity.

\section{Summary}

In summary, we have presented a detailed Cu $K$-edge RIXS study of
\lco~in which we resolve a multiplet of charge-transfer excitations
in the 1-7 eV range. We suggest several interpretations to explain
the polarization-dependent spectra. A calculation applying
third-order perturbation theory introduces a final-energy resonance
and successfully simulates the main characteristics of the spectra
for both polarization conditions. On the other hand, transitions to
continuum $4p$ bands that begin at the main absorption edge as well
as $4p_\sigma\rightarrow4p_z$ relaxation are offered as alternative
explanations to the fluorescence-like component in the contours.
These proposals all emphasize the important role of the $4p$
electrons in the RIXS cross section.

\vspace{5mm}
\section{Acknowledgements} The authors gratefully acknowledge valuable discussions
with P. Abbamonte, U. Bergmann, J. van den Brink, T. P.
Devereaux, M. V. Klein, Y. J. Kim, K.-W. Lee, R. S. Markiewicz, W. E.
Pickett, K. M. Shen, M. van Veenendaal, and F. C. Zhang. The work at Stanford University was supported by the DOE under Contract
No. DE-AC02-76SF00515 and by the NSF under Grant No. 0405655. Work at the CMC-XOR
Beamlines is supported in part by the Office of Basic Energy
Sciences of the U.S. Dept. of Energy and by the National Science
Foundation Division of Materials Research. Use of the Advanced
Photon Source is supported by the Office of Basic Energy Sciences of
the U.S. Department of Energy under Contract No. W-31-109-Eng-38

\bibliography{lcobib5_July10}

\end{document}